# Efficient free-space to chip coupling of ultrafast sub-ps THz pulse for biomolecule fingerprint sensing

Yanbing Qiu[1], Kun Meng[2], Wanling Wang[1], Jing Chen[1], John Cunningham[3], Ian Robertson[3], Binbin Hong[1,3,†], and Guo Ping Wang[1,*]

[1] *College of Electronics and Information Engineering, Shenzhen University, Shenzhen 518060, China*

[2] *Qingdao QUENDA Terahertz Technology Co. Ltd, Qingdao 266104, China*

[3] *School of Electronic and Electrical Engineering, University of Leeds, Leeds LS2 9JT, UK*

Corresponding authors: † E-mail: b.hong@szu.edu.cn; * E-mail: gpwang@szu.edu.cn.

**Abstract:**

Ultrafast sub-ps THz pulse conveys rich distinctive spectral fingerprints related to the vibrational or rotational modes of biomolecules and can be used to resolve the time-dependent dynamics of the motions. Thus, an efficient platform for enhancing the THz light-matter interaction is strongly demanded. Waveguides, owing to their tightly spatial confinement of the electromagnetic fields and the longer interaction distance, are promising platforms. However, the efficient feeding of the sub-ps THz pulse to the waveguides remains challenging due to the ultra-wide bandwidth property of the ultrafast signal. We propose a sensing chip comprised of a pair of back-to-back Vivaldi antennas and a 90° bent slotline waveguide to overcome the challenge. The effective operating bandwidth of the sensing chip ranges from 0.2 to 1.15 THz, with the free-space to chip coupling efficiency up to 50%. Over the entire band, the THz signal is 42.44 dB above the noise level with a peak of 73.40 dB. To take advantages of the efficient sensing chip, we have measured the characteristic fingerprint of α-lactose monohydrate, and a sharp absorption dip at near 0.53 THz has been successfully observed demonstrating the accuracy of the proposed solution. The proposed sensing chip has the advantages of efficient in-plane coupling, ultra-wide bandwidth, easy integration and fabrication, large-scale manufacturing capability, and cost-effective, and can be a strong candidate for THz light-matter interaction platform.

spectroscopy.

## 1. Introduction

Terahertz (THz) wave has characteristic scales in the dimensions of time, space, and photo energy that are closely related to many unique phenomena in the microscopic world, which results in various application fields ranging from spectroscopy [1], wireless communication [2], security imaging [3], non-destructive testing [4], and radio astronomy [5]. Sub-ps THz pulse conveying ultra-wideband spectral information can be used to resolve the time-dependent ultrafast dynamics of molecules which are associated with vibrational and rotational motions, and long-range intermolecular interactions [6, 7]. The Light-matter interactions between the ultrafast sub-ps THz pulses and the materials to be tested have recently been active research areas to reveal the physical mechanisms associated with the intrinsic properties of the materials, including spectral fingerprints, dielectric properties, carrier lifetimes, etc. [8]

Limited by the diffraction limit, the free-space THz beam is inefficient when interacting with nanomaterials with ultra-small quantity and volume, such as biomolecules, perovskite films, and two-dimensional materials. Waveguides have been widely demonstrated to be able to enhance the light-matter interaction due to the tightly spatial confinement of the electromagnetic fields and the longer interaction distance. In particular, plasmonic waveguides which confine the electromagnetic fields into deep-subwavelength regions and hence break the diffraction limit are very promising platforms for strong light-matter interactions [9]. However, there are a few challenges to realizing strong interaction between the sub-ps THz pulse and the nanomaterials on/in waveguides. To efficiently utilize the large bandwidth advantage of the sub-ps THz pulse, the waveguides conveying the signal have firstly to be able to support single-mode operation over decades of bandwidth. Many THz waveguides or fibers are tailored for low-loss wave propagation and do not support such wide operating bandwidth, including photonic crystal fibers [10 - 16], microstructure fibers [17 - 20], step-index fibers [21 - 23], substrate-integrated waveguides [24], silicon waveguides [25 - 27], etc. Some classic metallic waveguides, however, support ultra-wideband single-mode operation costing higher Ohmic losses, like, parallel-plate waveguide [28, 29], microstrip [30], coplanar waveguide [31], slotline [32], metal wires [33 - 35], Goubau line [36], etc. Considering that the THz light-matter interaction on/in waveguides does not require a long propagation distance and hence can tolerate slightly higher waveguide losses, the metallic waveguides are more suitable to be the interaction platform.

The other challenge Is the ways of coupling the ultrafast sub-ps THz pulse into the waveguides. One way is to in-situ excite and detect the THz pulse in the waveguides integrated with LT-GaAs based photoconductive switches [31, 36, 37 - 50]. This way has high integration level, small footprint, and high humid air immunity, but it also requires difficult fabrication processes as well as high fabrication and material costs. The other way is to couple the THz pulse into the waveguides using external THz sources. THz time-domain spectroscopy (THz-TDS) is commonly used in this scenario. D. Grischkowsky *et al.* experimentally reported a quasi-optical method to couple sub-ps THz pulse into submillimeter circular metal waveguides using hyper hemispherical silicon lenses in 1999 [51]. In the next year, they used the cylindrical lens to couple the sub-ps THz pulse into plastic ribbon waveguides [52]. Due to the large group-velocity dispersion of both circular metal waveguides and plastic ribbon waveguides, the received THz pulses are very dispersive in both of the above-mentioned works. In 2001, an ultra-wideband THz signal ranging from 0.1 to 4 THz is coupled into a parallel-plate waveguide using cylindrical dielectric lenses [28]. Since then, owing to the excellent performances in bandwidth, loss, and dispersion, parallel-plate waveguides have widely been studied [53, 54], and many guided THz spectroscopy applications based on the parallel-plate waveguide have been demonstrated to detect the fingerprints of water, biological molecules, etc. [53, 55] However, despite of their advantages, parallel-plate waveguides are relatively bulky and inflexible for bending and integration which limits its large-scale applications. In 2004, Mittleman et al. experimentally demonstrated the free-space to bare metal wires coupling of THz signal using focusing lenses and scattering input coupler, and the transmitted signal is dispersionless and has a bandwidth of about 0.1-0.5 THz [33]. In addition to time-domain methods, frequency-domain spectroscopy using electronic-based sources and detectors has also been studied. For example, Unlu et al. reported that an antenna array could be used to couple the continuous-wave THz wave into a spoof surface plasmon polariton waveguide [56]. Xie et al. proved an out-of-plane THz free-space to silicon waveguide coupling using a grating and a compact spot-size converter operating from 170 to 220 GHz with a coupling efficiency of about 5 dB at 194 GHz [57]. Usually, electronic-based frequency-domain spectroscopy (FDS) is relatively narrow-band and thus has rarely been used for fingerprint detection.

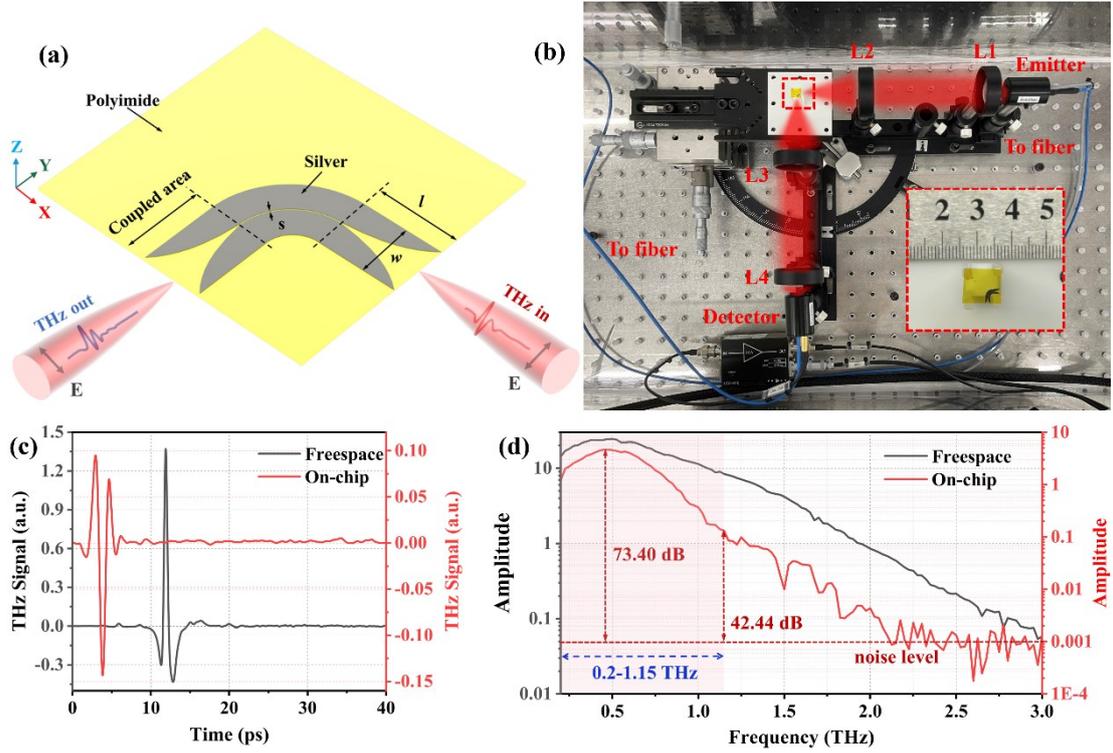

**Figure 1. Schematic, experimental set-up, and measurement results of the free-space to chip coupling system.** (a) Sensing chip schematic diagram. (b) Experimental set-up. L1, L2, L3, and L4 are TPX lenses. Emitter and Detector refer to the THz photoconductive antennas. The red semi-transparent region indicates the optical path of the THz signal. The inset diagram shows the fabricated sensing chip. (c) Free-space (black) and on-chip (red) time domain signals obtained using terahertz time-domain spectroscopy (THz-TDS) system; (d) Free-space (black) and on-chip (red) spectra.

Here, we numerically investigate and experimentally demonstrate an efficient in-plane free-space to chip coupling of ultrafast sub-ps THz pulse using ultra-wideband Vivaldi antennas. The sensing chip is comprised of a pair of back-to-back Vivaldi antennas and a 90° bent slotline waveguide. The sensing chip can easily be fabricated using photolithography and metal physical vapor deposition on a large scale with high consistency and low cost. Due to the flat form of the sensing chip and the in-plane coupling, the chip is highly compatible with commercial THz-TDS and FDS instruments. In addition, we also demonstrate the detection of the biomolecule vibration fingerprint by placing the α-lactose monohydrate powders on the 90° bent slotline waveguide region. The proposed THz sensing chip with an efficient in-plane coupling function can be a promising platform for ultrafast light-matter interaction.

## 2. The design and performances of the sensing chip

The schematic diagram of the proposed sensing chip manufactured on the polyimide (PI) is shown in Figure 1(a). Polyimide, with its excellent heat resistance, fine strength, rigidity, and compatibility with microfabrication processes, is one of the best organic polymer materials available with high comprehensive performance. Meanwhile, 13 μm thick film is chosen because it can suppress the competition from the high-order substrate mode to the fundamental slotline mode. The sensing chip consists of a Vivaldi antenna at each end and a section of 90° bent slotline waveguide. Vivaldi antennas are chosen due to their flat form, in-plane coupling, broad bandwidth, low cross-polarization, easy integration, and easy fabrication. The two back-to-back Vivaldi antennas have the same specifications and dimensions. The length and the opening width of the Vivaldi antenna are 1.6 mm ($\ell$) and 0.62 mm ($w$), respectively. Here, the top curvature of the right Vivaldi antenna follows the exponential function expressed as $y = A\left(e^{Bx} - 1\right) + \frac{s}{2}$, where A and B are tentative geometrical constants, and $s$ is the width of the waveguide. The original point is placed at the center of the joint interface between the 90° bent slotline waveguide and the right Vivaldi antenna.

The sensing chip is fabricated using the lift-off technique. The PI film firstly adhered to a silicon wafer, and then the negative photoresist was spun on the PI film. After that, the sensing chip pattern was transferred onto the PI film with a pre-fabricated photomask using photolithography. Whereafter, a 10 nm thick titanium layer, a 300 nm thick silver layer, and a 10 nm thick titanium layer were orderly deposited on top of the patterned polyimide substrate by electron-beam evaporation. The first titanium layer acts as an adhesive layer between the PI film and the silver layer, while the second titanium layer is a protective layer preventing the rapid oxidation of the silver layer. Finally, the sensing chip was successfully patterned on the PI film after the photoresist was dissolved in acetone and rinsed in DI-$H_2O$.

The width of the channel ($s$) in the curved slotline waveguide is set to be 25 μm, which is about one-twenty fourth of the representative wavelength at 500 GHz (the peak of the THz spectrum), for supporting broadband the single-mode operation. The slotline waveguide is intentionally designed to be 90° bent, so the polarizations of the THz signals are orthogonal to each other at the input and the output ends. This can eliminate unwanted scattering signals which are not coupled into the slotline waveguide, to make the received signal cleaner.

Figure 1(b) shows the experimental set-up for testing the proposed sensing chip, and the embedded graph shows the optical image of the fabricated sensing chip. A fiber-based THz-TDS system is used to focus the THz radiation into the slotline waveguides with the help of terahertz TPX lenses. As shown in the diagram, the terahertz signal emitted by the emitter is coupled into the sensing chip by the Vivaldi antenna after diverging at the L1 lens and converging at the L2 lens. After that, the THz signal is transmitted over a curved slotline waveguide and then coupled into free space by the second Vivaldi antenna. Finally, it passes through the L3 and L4 lenses and is received by the detector. The sensing chip attached to a 3D-printed holder is fixed to a three-dimensional adjustable stage helping the alignment of the optical path of the THz signal. The system is sealed in an acrylic glove box, where the effect of water absorption during measurements is eliminated by purging dry nitrogen. The terahertz time-domain signals shown in Figure 1(c) are measured using this experimental system. The unloaded time-domain signal passing through the chip (red curve) is obtained from the 90° bent testing system, as shown in Figure 1(b). While the time-domain signal of free space (black curve) is obtained from a testing system with a straight optical path. Figure 1(d) shows the THz amplitude spectra after doing the fast Fourier transform (FFT) to the measured time-domain signals shown in Figure 1(c). It can be seen that the effective operating frequency band of the sensing chip is 0.2-1.15 THz (red shaded area). Over this operating band, the overall terahertz signal is 42.44 dB higher than the noise level, and the highest dynamic range can reach 73.40 dB. As there is a strong dip that occurred at around 1.23 THz and 1.5 THz, which is caused by unwanted resonances on the chips, so we consider the effective operating band is slightly below this frequency point to obtain a relatively clean transmission window.

**3. Vivaldi antenna optimization for enhancing the coupling efficiency**

The directivity of the Vivaldi antenna is one of the key factors that affect the overall coupling efficiency between the free-space terahertz beam and the guided slotline mode. The Vivaldi antenna acts differently at the input and output ends. At the input end, it receives a concentrated THz beam with a good transverse Gaussian shape, while at the output end, its radiation patterns are shown in Figure 2. Figure 2 shows the far-field radiation patterns of five Vivaldi antennas with different lengths at several representative frequency points. The tentative geometrical constants *A* and *B* corresponding to the five antenna lengths are shown in Table 1. These far-field diagrams were simulated using the CST microwave studio. Time-Domain Solver is used to calculate the propagation field patterns of the slotline waveguide and the Vivaldi antenna. The frequency-dependent permittivity of the PI and the conductivity of the silver are obtained from the CST material library. The typical permittivity of the PI at 1 THz is $\varepsilon = \varepsilon' + i\varepsilon''$, where $\varepsilon'=3.4215$ and $\varepsilon''=0.0095$. And the typical conductivity of the silver is $6.3012\times10^7$ S/m. The boundary conditions in all directions are set to be open. For simplicity, surface roughness is not included in the simulation.

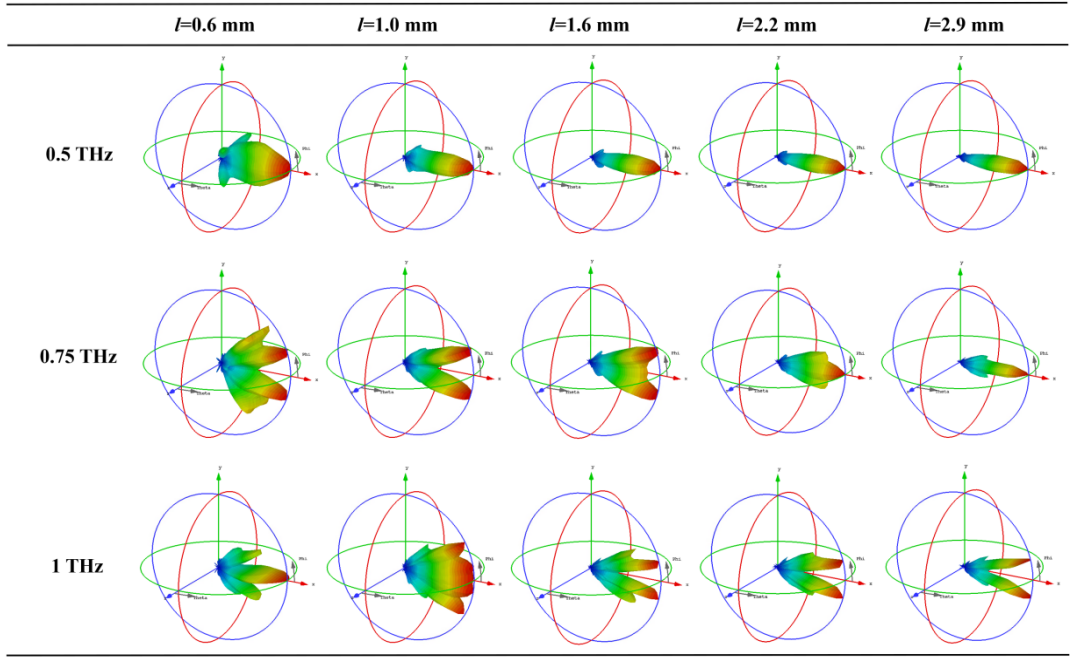

**Figure 2. Simulated far-field radiation patterns of the Vivaldi antennas with different lengths at several representative frequency points.** The widths of the Vivaldi antennas (w) and the channel of the slotline waveguide are all fixed as 0.62 mm and 25 μm, respectively.

Table 1. The tentative geometrical constants of the representative Vivaldi antennas with different lengths

|   | *l*=0.6 mm | *l*=1.0 mm | *l*=1.6 mm | *l*=2.2 mm | *l*=2.9 mm |
|---|---|---|---|---|---|
| *A* | 0.535 | 0.0375 | 0.026 | 0.015 | 0.0145 |
| *B* | 4.2 | 2.85 | 2 | 1.7 | 1.3 |

    Ideally, a Vivaldi antenna with high directivity over a wide band is preferred, but due to the limit of the intrinsic bandwidth of the Vivaldi antenna, the radiation patterns of the Vivaldi antennas diverge significantly towards high frequencies, as shown in Figure 2. Besides, it can be seen that the Vivaldi antenna with a longer antenna length has better directivity, but at the same time, it will also introduce

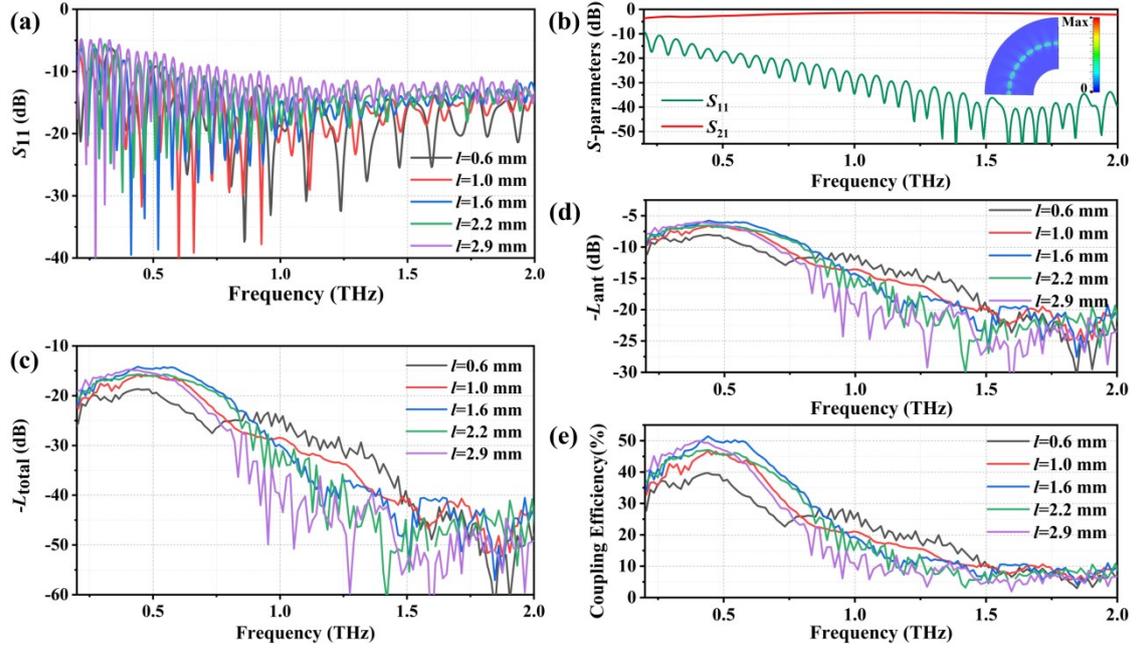

**Figure 3. Simulated and measured performances of the Vivaldi antennas and the slotline waveguide.** (a) and (b) are reflection coefficients of antennas with different lengths (a) and transmission coefficients and reflection coefficients of bent waveguide (b), which are obtained by using CST simulation. (c)-(e) are the total insertion losses of the sensing chip (c), the insertion loss of a single Vivaldi antenna (d), and the coupling efficiency of a single Vivaldi antenna (e), respectively, which are obtained by measurement.

more Ohmic loss. Thus, a balance between good directivity and low insertion loss has to be considered in the design. It should be noted that the focal length and the diameter of the L2 and L3 TPX lenses used to focus and collimate the THz signal are 50 mm and 38 mm, respectively, which indicates the maximum ray angle against the axis is $\theta_0$=20.8°. Therefore, the THz field radiated by the Vivaldi antenna at the output end with a radiation angle large than $\theta_0$ will not be picked up by the L3 lens. On the other hand, the L3 lens can pick up the THz field up to the largest angle of 20.8°, suggesting that the coupling scheme can tolerate the divergency of the Vivaldi antenna at high THz frequencies to a certain extent.

Along with the directivity, the reflection coefficient ($S_{11}$) also reveals the coupling property of the Vivaldi antenna, as shown in Figure 3(a). Here, Vivaldi antennas with different lengths and a fixed width are fed by a slotline waveguide with a fixed gap width of 25 μm. Comparing Figure 3(a) with Figure 2, although the directivity of the Vivaldi antenna with $l$=2.9 mm is best among the five selected representative lengths, its reflection coefficient is the largest. Generally, it can be seen from Figure 3(a) that, the Vivaldi antennas can support wideband operation as their reflection coefficient is low between 0.2 – 2 THz. At the low THz frequency end, the reflection coefficients are relatively large indicating that the impedance mismatch between the Vivaldi antenna using the proposed geometric properties and the long THz free-space wave is stronger. Apart from the case of *l*=2.9 mm, the reflection coefficients of other Vivaldi antennas are basically below -10 dB between 0.5 – 2 THz. Among the five selected representative Vivaldi antennas, the longer Vivaldi antenna shows better directivity, while the shorter antenna shows a lower reflection coefficient, and the overall coupling coefficient is affected by both factors. Therefore, the THz slotline chips with these five representative Vivaldi antennas are fabricated and measured to experimentally compare their overall free-space to chip coupling performance.

Figure 3(b) shows the experimentally measured total insertion losses of the chips ($L_{total}$) with different Vivaldi antennas. They are determined using the following equation:

$$L_{total} = -20 \times \log_{10}\left(\frac{A_{chip}}{A_{fs}}\right). \tag{1}$$

Here, $A_{fs}$ denotes the frequency domain amplitude of the free-space THz signal measured using a straight THz optical path without the chip being placed between the L2 and the L3 lenses, which has been illustrated using a black solid line in Figure 1(d). $A_{chip}$ denotes the amplitude of the on-chip THz signal measured using a 90° bent THz optical path with the chip being placed between the L2 and the L3 lenses, an example of which has been illustrated using a red solid line in Figure 1(d). It can be seen from Figure 3(b) that the overall performance of the chip with *l*=1.6 mm Vivaldi antenna is optimal. Averagely, it (blue curve) has the minimum total loss across the frequency band of interest ranging from 0.2 to 1.15 THz. In addition, the ripples, which are caused by impedance mismatch and unwanted resonances on the chip, are also minimal on the curve for *l*=1.6 mm.

Then, the insertion loss of a single Vivaldi antenna ($L_{ant}$) can be obtained using the following expression:

$$L_{ant} = \frac{L_{total} - L_{wg}}{2}, \tag{2}$$

where $L_{wg}$ is the insertion loss of the 90° bent slotline waveguide. It should be noted that $L_{wg}$ is obtained numerically based on CST simulation due to the lack of a robust way to extract its value experimentally. It is a rational approximation because the slotline waveguide with 25 μm gap width supports the wideband single-mode operation and the 90° bent structure is smooth and has no undesired resonance feature in it. The dispersion of the dielectric properties of the polyimide substrate and the conductivity of the silver layer are taken into account in the simulation using CST Microwave Studio. For simplification, the surface roughness of materials is not considered. The $L_{wg}$ is equivalent to the transmission coefficient ($S_{21}$) of the 90° bent slotline waveguide, as shown in Figure 3(c). It can be seen that the slotline waveguide supports wideband transmission between 0.2 THz and 2 THz, and its insertion loss ranges from 1.36 dB to 3.66 dB. Furthermore, the inset presents the normalized electric field propagating along the bent waveguide at 0.5 THz, which shows that the THz field is tightly confined around the gap of the slotline waveguide. The insertion losses of the Vivaldi antennas calculated using equation (2) are shown in Figure 3(d). The Vivaldi antenna with a length of 1.6 mm is still the optimal one, with the minimum loss of only 5.77 dB occurring at 0.44 THz.

Finally, the coupling efficiency of a single Vivaldi antenna on the chip can be calculated as follows:

$$\eta = 10^{\frac{-L_{ant}}{20}} \times 100\% . \qquad (3)$$

The calculated results are shown in Figure 3(e). It can be seen that the overall coupling efficiency of the antenna with a length of 1.6 mm can reach 50%, which is the most optimal among different antennas.

Figure 4 shows the radiation patterns of Vivaldi antennas with different widths at the output end. Their corresponding tentative geometrical constants are given in Table 2. It can be seen that the radiation directivities of the antennas gradually weaken at high frequencies. This property is similar to that of the antennas with different lengths discussed in Figure 2. Meanwhile, among the five selected antennas, the antennas with 0.45 mm and 0.62 mm opening widths show a relatively better balance between high directivity and low return loss, which will be further discussed in Figure 5(a). In contrast, antennas with smaller and larger widths exhibit relatively poor radiative directivity at 0.5 THz. At the same time, the reflection coefficients of the antennas have also been simulated using the CST microwave studio, as shown in Figure 5(a). Comparing Figure 4 with Figure 5(a), although the directivity of the Vivaldi antenna with $w$=0.28 mm is the best among the five selected representative antennas, its reflection coefficient is the largest. In addition, the reflectance coefficients of the antennas with widths of 0.62, 0.79, and 0.96 mm are very similar and are smaller than that of the antenna with w=0.45 mm. Eventually, two chips with antenna widths of 0.45 and 0.62 mm were fabricated and measured. The reason for this choice is that the antenna with a width of 0.45 mm has excellent radiative directivity, while the antenna with w=0.62 mm shows a better reflection coefficient. It is difficult to further compare the potential performances of the chips with the antennas width of 0.45 and 0.62 mm merely using the simulation results, as both the directivity and the reflection coefficient affect the overall coupling efficiency and a tradeoff has to be made between them. Thus, we have fabricated and measured the two chips to experimentally determine which one performs better.

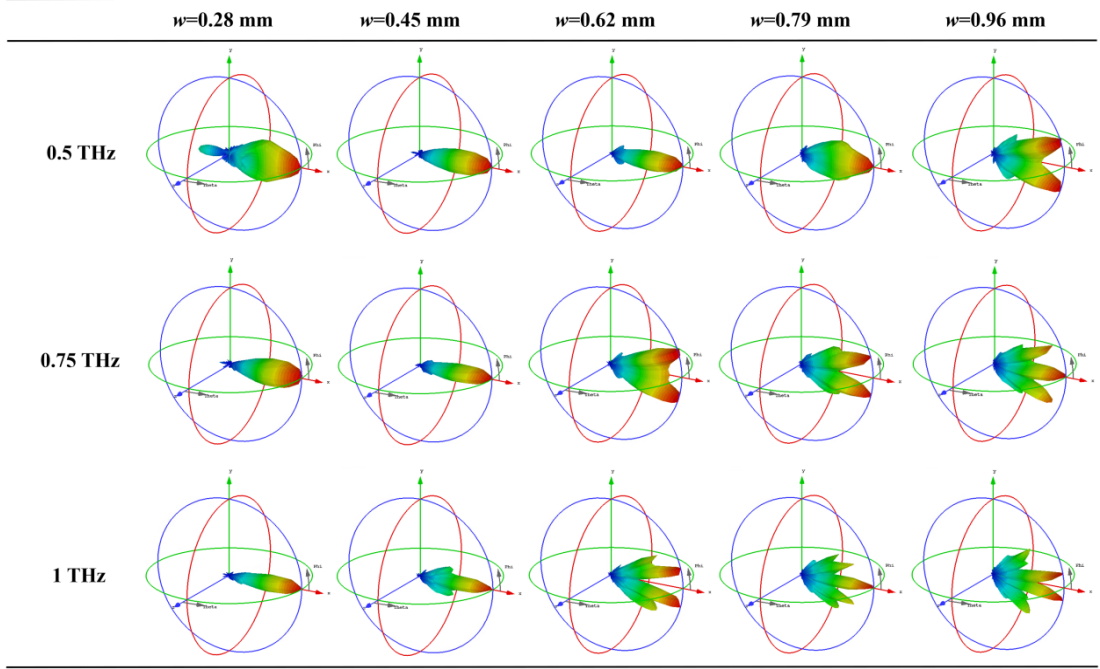

**Figure 4. Simulated far-field radiation patterns of the Vivaldi antennas with different widths at several representative frequency points.** The length of the Vivaldi antennas (l) and the central channel width (s) of the slotline waveguide are fixed as 1.6 mm and 25 μm, respectively.

Table 2. The tentative geometrical constants of the representative Vivaldi antennas with different widths

|   | w=0.28 mm | w=0.45 mm | w=0.62 mm | w=0.79 mm | w=0.96 mm |
| --- | --- | --- | --- | --- | --- |
| A | 0.026 | 0.026 | 0.026 | 0.026 | 0.026 |
| B | 1.52 | 1.8 | 2 | 2.15 | 2.265 |

Figure 5(b) shows the experimentally measured $-L_{total}$ of the sensing chip with different Vivaldi antennas, calculated using the equation (1). It can be seen that the case with w=0.62 mm has a lower total loss than the other case. Besides, at high frequencies near 1 THz, the spectrum of the former case is smoother than the latter one, which is important for sensing applications. Figure 5(c) presents the insertion losses of the one-side Vivaldi antennas on the chips, obtained using equation (2). Here, same as Figure 3(d), the insertion loss of the 90°bent slotline waveguide used to calculate the $L_{ant}$ is given in Figure 3(c). The Vivaldi antenna with $w$=0.62 mm shows relatively lower insertion loss over a wide frequency band. Finally, the corresponding coupling efficiencies of the two Vivaldi antennas are shown in Figure 5(d). It can be concluded from Figures 5(b) -5(d) that the Vivaldi antenna with $w$=0.62 mm is superior to the other antenna with $w$=0.45 mm in the aspects of the total loss, single antenna loss, spectrum smoothness, and coupling efficiency. Hence, the former design has been chosen for THz fingerprint sensing the biomolecule, which will be discussed in Figure 6.

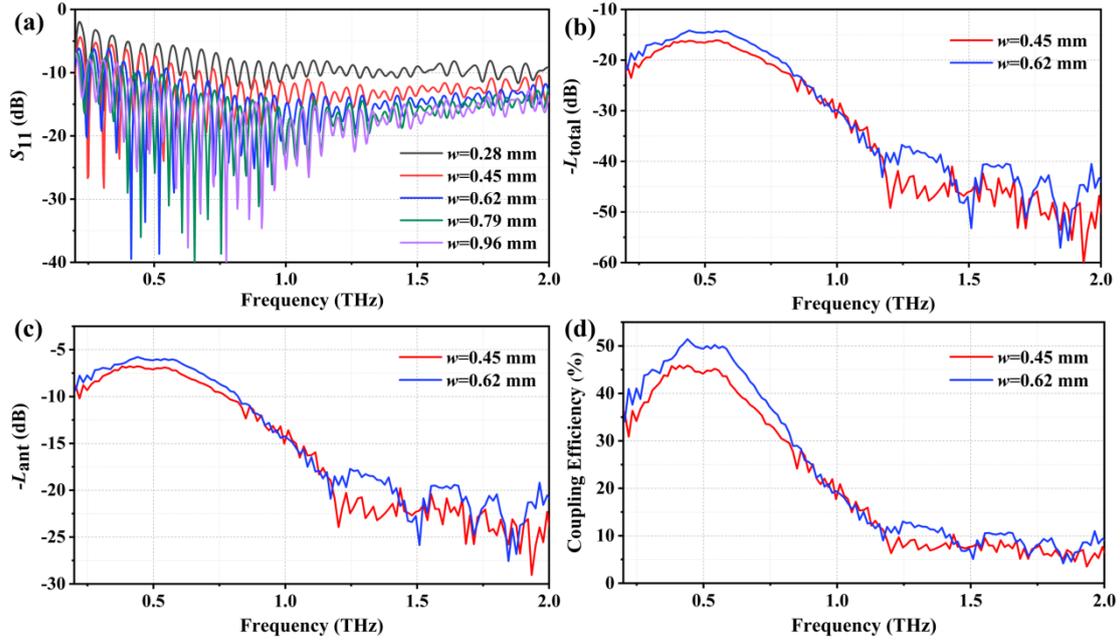

**Figure 5. Simulated and measured performances of the Vivaldi antennas with different widths.** (a) is reflection coefficients of antennas, which are obtained by using CST simulation. (b)-(d) are the total insertion losses of the sensing chip (b), the insertion loss of a single Vivaldi antenna (c), and the coupling efficiency of a single Vivaldi antenna (d), respectively, which are obtained by measurement.

## 4. On-chip biomolecule vibration fingerprint sensing

To take advantages of the sensing chip with efficient free-space to chip coupling, we investigate the light-matter interaction using it. The THz spectrum is abundant in distinctive spectral fingerprints relating to the vibration and rotation of most biomolecules, which has wide application potential in biomedical sensing and imaging [58]. The α-lactose monohydrate is a widely studied biomolecule that has a THz absorption fingerprint at around 0.53 THz due to an external hindered rotational mode along the B-axis of α-lactose crystal with intermolecular hydrogen-bond networks [59, 60]. Therefore, the α-lactose monohydrate is an ideal candidate to demonstrate and calibrate the fingerprint sensing ability of the proposed sensing chip.

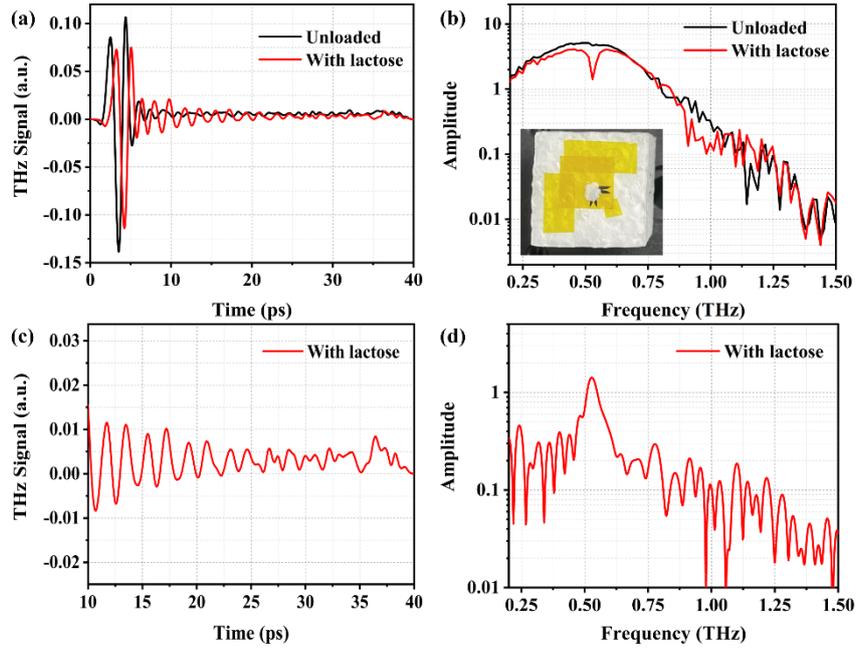

**Figure 6. The THz signals and spectra with and without the α-lactose monohydrate powders on the chip.** (a)The measured time-domain signals with and without (unloaded) the α-lactose monohydrate powders; (b) Frequency spectra related to the signals shown in (a). The inset diagram shows the sensing chip with α-lactose monohydrate powder added. The mass of the monohydrate powder is 1.0 mg. The sensing chip is placed on the top of an EPS foam; (c) Truncated signal with lactose extracted from (a); (b) The frequency spectrum of (c).

The interaction between the ultrafast THz pulse and the α-lactose monohydrate happens on the slotline waveguide region of the sensing chip, where the electromagnetic field is tightly confined in the deep-subwavelength scale enhancing the interaction efficiency. Figure 6 shows the measured time-resolved signals and their spectra calculated using FFT (Fast Fourier Transform) of the above light-matter interaction processes. Figure 6(a) compares the time-domain signals of the THz waves passing through the sensing chip with (red curve, loaded chip) and without (black curve, unloaded chip) 1.0 mg α-lactose monohydrate powders. The time step is 33.4 fs. The signal with lactose on the chip is delayed and attenuated compared with the unloaded one, and it is more fluctuant after the main peaks.

The fluctuant is mainly attributed to the rotation of the α-lactose crystal.

Figure 6(b) presents the frequency spectra calculated based on Figure 6(a) using FFT. The inset shows the optical photo of the chip with monohydrate powder placed on low-density expanded polystyrene (EPS) foam. The EPS foam is dispersionless and low-loss. Over the frequency range of interest from 0.2 – 2 THz, its refractive index is very close to that of air and is measured to be 1.006 with negligible variations, and its absorption is extremely low and is measured to be less than 0.35 cm$^{-1}$. Thus, the supporting EPS foam has negligible impacts on the transmission properties of the proposed chip. Besides, the EPS foam has good mechanical stability which can support the polyimide-based chip firmly when adding the monohydrate samples. In the figure, a sharp absorption dip near 0.53 THz can be observed from the red curve, while it does not appear on the black curve, indicating that the proposed chip can efficiently sense the THz fingerprint of the monohydrate with a small quantity. At frequencies above 0.75 THz, the spectra of unloaded and loaded chips are both oscillating due to the unwanted impedance mismatches occurring on the chip and also in the free-space optical system, which limits the useful bandwidth of the proposed chip.

Figure 6(a) and 6(d) show the truncated time-domain signal between 10 to 40 ps extracted from Figure 6(a) and its corresponding spectrum. The main peak in the frequency spectrum occurs at near 0.53 THz, indicating the damped fluctuation in Figure 6(a) is mainly caused by the hindered rotational mode of the α-lactose crystal. Consequently, we have successfully experimentally demonstrated a wideband and efficient method to excite the THz deep-subwavelength plasmonic mode of the slotline waveguide via the Vivaldi antenna, which can be used as a platform for detecting the THz fingerprints of biomolecules on a chip. Further efforts can be made to broaden the smooth spectrum of the chip in future works.

## 5. Conclusion

In summary, we proposed a sensing chip that is comprised of a pair of back-to-back Vivaldi antennas and a 90° bent slotline waveguide, to overcome the challenge of efficient free-space to chip coupling as well as enhanced light-matter interaction on the chip. The coupling efficiency and fingerprint sensing ability of the proposed chip have been numerically investigated and experimentally demonstrated. With the help of Vivaldi antennas, the in-plane free-space to chip coupling efficiency is up to 50% in the operating frequency range from 0.2 to 1.15 THz. The amplitude of the THz signal is 42.44 dB above the noise level with a peak of 73.40 dB across the entire operating band. On-chip fingerprint sensing of α-lactose monohydrate has been performed and an absorption dip at near 0.53 THz has been observed. The proposed sensing chip with efficient free-space to chip coupling can be a promising light-matter inaction platform between the ultrafast sub-ps THz pulse and many nanomaterials.

## Acknowledgments

This work was supported by the National Natural Science Foundation of China (Grant Nos. 62105213, 12074267, and 11734012), the Guangdong Basic and Applied Basic Research Fund (Grant No. 2020A1515111037, and 2021A1515011713), the Key Areas R&D Program of Guangdong Province (Grant No. 2020B010190001), and the Shenzhen Fundamental Research Program (Grant No. 20200814113625003, and 20200813224730001). The authors thank Shuting Fang and Xudong Liu for their help in the experiments.

## Conflict of interest

The authors declare no conflict of interest.

## References

[1] P. U. Jepsen, D. G. Cooke, and M. Koch, "Terahertz spectroscopy and imaging–Modern techniques and applications," Laser & Photonics Reviews **5**(1), 124–166 (2011)

[2] H. J. Song and T. Nagatsuma, "Present and future of terahertz communications," IEEE Transactions on Terahertz Science and Technology **1**(1), 256–263 (2011).

[3] S. S. Dhillon, M. S. Vitiello, E. H. Linfield, A. G. Davis, M. C. Hoffmann, and J. Booske, "The 2017 terahertz science and technology roadmap," Journal of Physics D: Applied Physics **50**(4), 043001 (2017).

[4] D. M. Mittleman, "Twenty years of terahertz imaging," Optics Express **26**(8), 9417–9431 (2018).

[5] C. Kulesa, "Terahertz spectroscopy for astronomy: From comets to cosmology," IEEE Transactions on Terahertz Science and Technology **1**(1), 232–240 (2011).

[6] L. Ho, M. Pepper, and P. Taday, "Signatures and fingerprints," Nature Photonics **2**(9), 541-543 (2008).

[7] A. G. Davies, A. D. Burnett, W. Fan, E. H. Linfield, and J. E. Cunningham, "Terahertz spectroscopy of explosives and drugs," Materials Today **11**(3), 18-26 (2008).

[8] D. Nicoletti and A. Cavalleri, "Nonlinear light–matter interaction at terahertz frequencies," Advances in Optics and Photonics **8**(3), 401-464 (2016).

[9] X. Q. Zhang, Q. Xu, L. B. Xia, Y. F. Li, J. Q. Gu, Z. Tian, C. M. Ouyang, J. G. Han and W. L. Zhang, "Terahertz surface plasmonic waves: a review," Advanced Photonics **2**(1), 014001(2020).

[10] A. Dupuis, K. Stoeffler, B. Ung, C. Dubois and M. Skorobogatiy, "Transmission measurements of hollow-core THz Bragg fibers" Journal of the Optical Society of America B **28**(4), 896–907 (2011).

[11] B. Hong, M. Swithenbank, N. Somjit, J. Cunningham and I. Robertson, "Asymptotically single-mode small-core terahertz Bragg fibre with low loss and low dispersion," Journal of Physics D: Applied Physics **50**(4), 045104 (2016).

[12] B. Hong, M. Swithenbank, N. Greenall, R. G. Clarke, N. Chudpooti, P. Akkaraekthalin, N. Somjit, J. E. Cunningham and I. D. Robertson, "Low-loss asymptotically singlemode THz Bragg fiber fabricated by digital light processing rapid prototyping," IEEE Transactions on Terahertz Science and Technology **8**(1), 90–99 (2018).

[13] B. Hong, N. Chudpooti, P. Akkaraekthalin, N. Somjit, J. Cunningham and I. Robertson, "Investigation of electromagnetic mode transition and filtering of an asymptotically single-mode hollow THz Bragg fibre," Journal of Physics D: Applied Physics **51**(30), 305101 (2018).

[14] J. Li, K. Nallappan, H. Guerboukha and M. Skorobogatiy, "3D printed hollow core terahertz Bragg waveguides with defect layers for surface sensing applications," Optics Express **25**(4), 4126–44 (2017).

[15] Z. Wu, W. R. Ng, M. E. Gehm and H. Xin, "Terahertz electromagnetic crystal waveguide fabricated by polymer jetting rapid prototyping," Optics Express **19**(5), 3962–72 (2011).

[16] J. Yang, J. Zhao, C. Gong, H. Tian, L. Sun, P. Chen, L. Lin and W. Liu, "3D printed low-loss THz waveguide based on Kagome photonic crystal structure," Optics Express **24**(20), 22454–60 (2016).

[17] V. Setti, L. Vincetti and A. Argyros, "Flexible tube lattice fibers for terahertz applications," Optics Express **21**(3), 3388–99 (2013).

[18] W. Lu, S. Lou and A. Argyros, "Investigation of flexible lowloss hollow-core fibres with tube-lattice cladding for terahertz radiation," IEEE Journal of Selected Topics in Quantum Electronics **22**(2), 214–20 (2016).

[19] K. Nielsen, "Bendable, low-loss Topas fibers for the terahertzfrequency range," Optics Express **17**(10), 8592–8601 (2009).

[20] J. Anthony, R. Leonhardt, A. Argyros and M. C. Large, "Characterization of a microstructured Zeonex terahertz fiber," Journal of the Optical Society of America B **28**(5), 1013–1018 (2011).

[21] K. Nallappan, Y. Cao, G. Xu, H. Guerboukha, C. Nerguizian and M. Skorobogatiy, "Dispersion-limited versus power-limited terahertz communication links using solid core subwavelength dielectric fibers," Photonics Research **8**(11), 1757-1775 (2020).

[22] H. Guerboukha, G. Yan, O. Skorobogata, and M. Skorobogatiy, "Silk foam terahertz waveguides," Advanced Optical Materials **2**(12), 1181-1192 (2014).

[23] B. Hong, Y. Qiu, N. Somjit, J. Cunningham, I. Robertson and G. P. Wang, "Guidance of Terahertz Wave over Commercial Optical Fiber," In 2021 46th International Conference on Infrared, Millimeter and Terahertz Waves (IRMMW-THz) (pp. 1-2). IEEE (2021).

[24] F. Fesharaki, T. Djerafi, M. Chaker, and K. Wu, "Guided-wave properties of mode-selective transmission line," IEEE Access **6**, 5379–5392 (2018).

[25] K. Tsuruda, M. Fujita, and T. Nagatsuma, "Extremely low-loss teraertz waveguide based on silicon photonic-

crystal slab," Optics Express **23**(25), 31977–31990 (2015).

[26] H. Amarloo and S. Safavi-Naeini, "Terahertz Line Defect Waveguide Based on Silicon-on-Glass Technology," IEEE Transactions on Terahertz Science and Technology **7**(4), 433–439 (2017).

[27] W. Gao, X. Yu, M. Fujita, T. Nagatsuma, C. Fumeaux, and W. Withayachumnankul, "Effective-medium-cladded dielectric waveguides for terahertz waves," Optics Express **27**(26), 38721–38734 (2019).

[28] R. Mendis and D. Grischkowsky, "Undistorted guided-wave propagation of subpicosecond terahertz pulses," Optics Letters **26**(11), 846-848 (2001).

[29] R. Mendis and D. Grischkowsky, "THz interconnect with low-loss and low-group velocity dispersion," IEEE Microwave and Wireless Components Letters **11**(11), 444-446 (2001).

[30] H. M. Heiliger, M. Nagel, H. G. Roskos, H. Kurz, F. Schnieder, W. Heinrich and K. Ploog, "Low-dispersion thin-film microstrip lines with cyclotene (benzocyclobutene) as dielectric medium," Applied Physics Letters **70**(17), 2233-2235 (1997).

[31] M. B. Ketchen, D. Grischkowsky, T. C. Chen, C. C. Chi, I. N. Duling Iii, N. J. Halas and G. P. Li, "Generation of subpicosecond electrical pulses on coplanar transmission lines," Applied Physics Letters **48**(12), 751-753 (1986).

[32] H. Pahlevaninezhad, B. Heshmat and T. E. Darcie, "Efficient terahertz slot-line waveguides," Optics Express **19**(26), B47-B55 (2011).

[33] K. Wang and D. M. Mittleman, "Metal wires for terahertz wave guiding," Nature **432**(7015), 376-379 (2004).

[34] T. I. Jeon, J. Zhang and D. J. Grischkowsky, "THz Sommerfeld wave propagation on a single metal wire," Applied Physics Letters **86**(16), 161904 (2005).

[35] A. Markov, H. Guerboukha and M. Skorobogatiy, "Hybrid metal wire–dielectric terahertz waveguides: challenges and opportunities," Journal of the Optical Society of America B **31**(11), 2587-2600 (2014).

[36] C. Russell, C. D. Wood, A. D. Burnett, L. Li, E. H. Linfield, A. G. Davies and J. E. Cunningham, "Spectroscopy of polycrystalline materials using thinned-substrate planar Goubau line at cryogenic temperatures," Lab on a Chip **13**(20), 4065-4070 (2013).

[37] D. R. Grischkowsky, "Optoelectronic characterization of transmission lines and waveguides by terahertz time-domain spectroscopy," IEEE Journal of Selected Topics in Quantum Electronics **6**(6), 1122-1135 (2000).

[38] M. B. Byrne, J. Cunningham, K. Tych, A. D. Burnett, M. R. Stringer, C. D. Wood, L. Dazhang, M. Lachab, E. H. Linfield, A. G. Davies, "Terahertz vibrational absorption spectroscopy using microstrip-line waveguides," Applied Physics Letters **93**(18), 182904 (2008).

[39] S. Yanagi, M. Onuma, J. Kitagawa and Y. Kadoya, "Propagation of terahertz pulses on coplanar strip-lines on low permittivity substrates and a spectroscopy application," Applied Physics Express **1**(1), 012009 (2008).

[40] S. Kasai, A. Tanabashi, K. Kajiki, T. Itsuji, R. Kurosaka, H. Yoneyama, M. Yamashita, H. Ito and T. Ouchi, "Micro strip line-based on-chip terahertz integrated devices for high sensitivity biosensors," Applied Physics Express **2**(6), 062401 (2009).

[41] J. Kitagawa, Y. Kadoya, K. Suekuni, M. A. Avila and T. Takabatake, "Characterization of terahertz absorption in clathrate compound by compact spectroscopic chip," Japanese Journal of Applied Physics **52**(2R), 028003 (2013).

[42] J. B. Wu, O. Sydoruk, A. S. Mayorov, C. D. Wood, D. Mistry, L. H. Li, E. H. Linfield, A. G. Davies and J. E. Cunningham, "Time-domain measurement of terahertz frequency magnetoplasmon resonances in a two-dimensional

electron system by the direct injection of picosecond pulsed currents," Applied Physics Letters **108**(9), 091109 (2016).

[43] C. Russell; M. Swithenbank; C. D. Wood; A. D. Burnett; L. H. Li; E. H. Linfield; A. G. Davies and J. E. Cunningham, "Integrated on-chip THz sensors for fluidic systems fabricated using flexible polyimide films," IEEE Transactions on Terahertz Science and Technology **6**(4), 619-624 (2016).

[44] J. Zuo, L. L. Zhang, C. Gong and C. L. Zhang, "Research progress of super-continuum terahertz source based on nano-structures and terahertz lab on-chip system," Acta Physica Sinica **65**(1), 010704 (2016).

[45] M. Swithenbank, A. D. Burnett, C. Russell, L. H. Li, A. G. Davies, E. H. Linfield, J. E. Cunningham and C. D. Wood, "On-chip terahertz-frequency measurements of liquids," Analytical Chemistry **89**(15), 7981-7987 (2017).

[46] R. Smith, A. Jooshesh, J. Y. Zhang and T. Darcie, "Photoconductive generation and detection of THz-bandwidth pulses using near-field coupling to a free-space metallic slit waveguide," Optics Express **25**(22), 26492-26499 (2017).

[47] P. Gallagher, C. S. Yang, T. Lyu, F. L. Tian, R. Kou, H. Zhang, K. Watanabe, T. Taniguchi and F. Wang, "Quantum-critical conductivity of the Dirac fluid in graphene," Science **364**(6436), 158-162 (2019).

[48] J. W. McIver, B. Schulte, F. U. Stein, T. Matsuyama, G. Jotzu, G. Meier and A. Cavalleri, "Light-induced anomalous Hall effect in graphene," Nature physics **16**(1), 38-41 (2020).

[49] J. O. Island, P. Kissin, J. Schalch, X. M. Cui, S. R. Ul Haque, A. Potts, T. Taniguchi, K. Watanabe, R. D. Averitt and A. F. Young, "On-chip terahertz modulation and emission with integrated graphene junctions," Applied Physics Letters **116**(16), 161104 (2020).

[50] K. Yoshioka, N. Kumada, K. Muraki and M. Hashisaka, "On-chip coherent frequency-domain THz spectroscopy for electrical transport," Applied Physics Letters **117**(16), 161103 (2020).

[51] R. W. McGowan, G. Gallot and D. Grischkowsky, "Propagation of ultrawideband short pulses of terahertz radiation through submillimeter-diameter circular waveguides," Optics Letters **24**(20), 1431-1433 (1999).

[52] R. Mendis and D. Grischkowsky, "Plastic ribbon THz waveguides," Journal of Applied Physics **88**(7), 4449-4451 (2000).

[53] J. Zhang and D. Grischkowsky, "Waveguide terahertz time-domain spectroscopy of nanometer water layers," Optics letters **29**(14), 1617-1619 (2004).

[54] H. Zhan, R. Mendis and D. M. Mittleman, "Superfocusing terahertz waves below λ/250 using plasmonic parallel-plate waveguides," Optics Express **18**(9), 9643-9650 (2010).

[55] N. Laman, S. S. Harsha, D. Grischkowsky and J. S. Melinger, "High-resolution waveguide THz spectroscopy of biological molecules," Biophysical journal **94**(3), 1010-1020 (2008).

[56] M. A. Unutmaz and M. Unlu, "Spoof surface plasmon polariton delay lines for terahertz phase shifters," Journal of Lightwave Technology **39**(10), 3187-3192 (2021).

[57] H. Zhang, C. Liang, J. Song, C. Fu, X. Zang, L. Chen and J. Xie, "Terahertz out-of-plane coupler based on compact spot-size converter," Chinese Optics Letters **20**(2), 021301 (2022).

[58] Z. Yan, L. G. Zhu, K. Meng, W. Huang and Q. Shi, "THz medical imaging: from in vitro to in vivo," Trends in Biotechnology **49**(7), 816-830 (2022).

[59] D. G. Allis, A. M. Fedor, T. M. Korter, J. E. Bjarnason and E. R. Brown, "Assignment of the lowest-lying THz absorption signatures in biotin and lactose monohydrate by solid-state density functional theory," Chemical Physics Letters **440**(4-6), 203-209 (2007).

[60] K. Moon, Y. Do, H. Park, J. Kim, H. Kang, G. Lee and H. Han, "Computed terahertz near-field mapping of molecular resonances of lactose stereo-isomer impurities with sub-attomole sensitivity," Scientific reports **9**(1), 1-8 (2019).